\def   \ni {\noindent}

\def   \ssk {\vskip  5truept}

\def   \bsk {\vskip 15truept}

\def   \newline {\hfil\break}

\def \sss {\scriptscriptstyle}
\def \ftn {\footnotesize}

\def\ergs{~{\rm erg}~{\rm s}^{-1} }
\def\ergscm2{~{\rm erg}~{\rm s}^{-1}~{\rm cm}^{-2} }
\def\MeV{~\rm{MeV}}
\def\keV{~\rm{keV}}

\def\G{~\rm{G}}
\def\s{~\rm{s}}

\def\Lg{$L_\gamma$}
\def\Lx{$L_{\sss X}$}
\def\edot{$L_{\rm sd}$}
\newbox\grsign \setbox\grsign=\hbox{$>$} \newdimen\grdimen \grdimen=\ht\grsign
\newbox\simlessbox \newbox\simgreatbox \newbox\simpropbox
\setbox\simgreatbox=\hbox{\raise.5ex\hbox{$>$}\llap
     {\lower.5ex\hbox{$\sim$}}}\ht1=\grdimen\dp1=0pt
\setbox\simlessbox=\hbox{\raise.5ex\hbox{$<$}\llap
     {\lower.5ex\hbox{$\sim$}}}\ht2=\grdimen\dp2=0pt
\setbox\simpropbox=\hbox{\raise.5ex\hbox{$\propto$}\llap
     {\lower.5ex\hbox{$\sim$}}}\ht2=\grdimen\dp2=0pt

\def \lsim {\hbox{\hspace{0.5mm}}\raise2pt
            \vbox{\moveleft0pt\hbox{$<$}}\lower2pt
            \vbox{\moveleft8pt\hbox{$\sim$ }}\hbox{\hskip 0.05mm}}

\def \t    {\thinspace}

\documentstyle[epsfig]{article}
\begin{document}


%

\hsize 5truein
\vsize 8truein
\font\abstract=cmr8
\font\keywords=cmr8
\font\caption=cmr8
\font\references=cmr8
\font\text=cmr10
\font\affiliation=cmssi10
\font\author=cmss10
\font\mc=cmss8
\font\title=cmssbx10 scaled\magstep2
\font\alcit=cmti7 scaled\magstephalf
\font\alcin=cmr6 
\font\ita=cmti8
\font\mma=cmr8
\def\ref{\par\noindent\hangindent 15pt}
\null


\title{\ni X-RAY AND GAMMA-RAY EMISSION \\ FROM PULSAR MAGNETOSPHERES}

\bsk \bsk
\author{\ni J.~Dyks, B.~Rudak}                                                       
\bsk
\ni \affiliation{Nicolaus Copernicus Astronomical Center, Rabia\'nska 8, 87-100 
Toru\'n, Poland}                                                
\bsk
\baselineskip = 12pt

\abstract{\ftn ABSTRACT 
\hskip 3mm For polar-cap models based on electromagnetic cascades induced
by curvature radiation of beam particles we calculate broad-band high-energy spectra of 
pulsed emission expected for classical ($\sim 10^{12}\G$) and 
millisecond pulsars ($\sim 10^9\G$). 
The spectra are a superposition of curvature and synchrotron 
radiation components and most of their detailed features depend significantly
on the magnetic field strength. 
The relations of expected pulsed luminosity \Lx~ (between 0.1~keV and 10~keV) 
as well as \Lg~ (above 100~keV) to
the spin-down luminosity \edot~ are presented.
We conclude that 
spectral properties and fluxes of pulsed non-thermal X-ray
emission of some objects, like the Crab or
the millisecond pulsar B1821$-$24, pose a challenge to 
the polar-cap models based on curvature and 
synchrotron radiation alone. On the other hand, such models may offer
an explanation
for the case of B1706$-$44.

}                                                    
\bsk
\baselineskip = 12pt

\bsk
\baselineskip = 12pt


\text{\ni 1. INTRODUCTION
\ssk     

The aim of this paper is to present general features of broad-band $X\gamma$ 
spectra expected for polar cap models with $X\gamma$ emission 
due to curvature (CR) and synchrotron (SR) processes
(e.g. Daugherty \& Harding 1982, Daugherty \& Harding 1996) and to show that
the luminosity of pulsed X-rays \Lx~ inferred for some pulsars is too high when
compared to their gamma-ray luminosity \Lg~ (or to their 
spin-down luminosity \edot~ when there
is no information about gamma-rays) to be understood within
polar-cap models unless some unorthodox assumptions are accepted.
Some of these features are only weakly model-dependent and may, therefore, play
a decisive role in assessing validity of polar-cap models. Details of the
results presented below are given by Rudak \& Dyks (1998b).

\bsk
\ni 2. SPECTRAL PROPERTIES OF CURVATURE RADIATION
\ssk 

For a monoenergetic injection rate function $Q_{\rm e}$ of beam particles cooled via CR,
their steady-state energy distribution (of a form $N_{\rm e} 
\propto \gamma^{-4}$) ranges from an injection energy $\gamma_0$ 
down to some lower limit $\gamma_{\rm break}$,
determined by the condition, that a cooling time-scale due to CR, $t_{\rm
cr}$,
is shorter 
than a time-scale for the particle to reside in the region of efficient CR cooling,
$t_{\rm esc}$. 
The corresponding photon spectrum of CR has a form: $N_{\rm cr}(\epsilon) 
\propto \epsilon^{-{5 \over 3}}$ and extends from $\epsilon_{\rm
cr}(\gamma_0)$ down to $\epsilon_{\rm break} \equiv \epsilon_{\rm 
cr}(\gamma_{\rm break})$ where $\epsilon_{\rm cr}(\gamma) = 1.5\, c\, \hbar\, \gamma^3 
\rho_{\rm cr}^{-2}$ ($\rho_{\rm cr}$ is a radius of local curvature).
Below $\epsilon_{\rm break}$ the photon spectrum follows the low energy tail 
of CR: $N_{\rm cr}(\epsilon) \propto \epsilon^{-{2 \over 3}}$.

\begin{figure}[t]
\centerline{\psfig{file=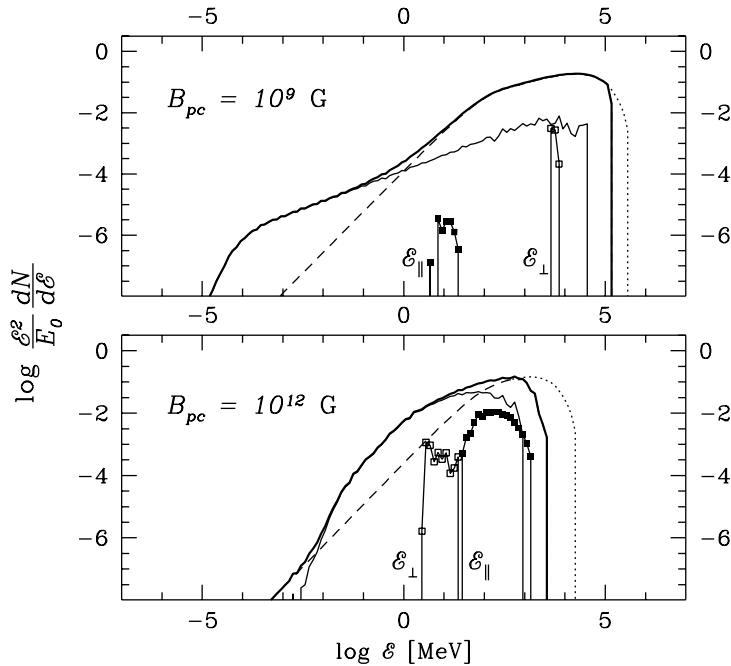, width=8.5cm}}
\caption{\ftn FIGURE 1. The radiation energy spectrum per logarithmic
bandwidth (thick solid line). 
The spectrum is normalized to the energy of the 
parent particle, $E_0$. 
Dashed line represents its CR component, part of which (dotted line) has been 
reprocessed 
by magnetic absorption into e$^\pm$-pairs. Thin solid line is the SR component.
In addition to the electromagnetic spectra, we show the spectra of
e$^\pm$-pairs: the line connecting filled squares is for the energy parallel 
to local magnetic field lines (${\cal E}_\parallel = \gamma_\parallel \, m_{\rm e} c^2$);
the line connecting open squares is the initial distribution of energy 
perpendicular to local magnetic lines (${\cal E}_\perp = \gamma_\perp \, m_{\rm e} c^2$).
The upper panel is for $B_{\rm pc}=10^{9}\G$,
$P = 3.1\times 10^{-3}\s$ 
(i.e. $L_{\rm sd} = 10^{35}\ergs$) and 
$E_0 = 1.08 \times 10^7 \MeV$.
The lower panel is for $B_{\rm pc}=10^{12}\G$, $P = 5.6\times 10^{-2}\s$
(i.e. $L_{\rm sd} = 10^{36}\ergs$) and
$E_0 = 6.68 \times 10^6 \MeV$.
}
\end{figure}


To find $\gamma_{\rm break}$ analytically,
we can approximate $t_{\rm esc}$ with $\rho_{\rm cr}/ c$
and from the 
condition $t_{\rm cr} = t_{\rm esc}$
we obtain the photon energy at which the spectral break 
should occur:
$\epsilon_{\rm break} = {9 \over 4} \, \hbar \, e^{-2} \, m_{\rm e} c^3 \approx 150$ MeV.
It does not depend on any pulsar parameters (in particular, on magnetic field
structure) as long as our estimate of $t_{\rm esc}$ is accurate. 

The CR component, 
dominating the gamma-ray domain, is visible in both panels of Fig.1
which show
numerically calculated spectra of both CR and SR
(and with transfer effects due to magnetic absorption included) per beam 
particle of initial energy $E_0$ (Rudak \& Dyks 1998a), 
injected at the polar cap's rim of a pulsar with a dipolar magnetic field 
of $B_{\rm pc}=10^{9}\G$ (upper panel), and $B_{\rm pc}=10^{12}\G$ (lower panel).
The spectral break at around $100\MeV$ is particularly pronounced in the former case.

\bsk
\ni 3. SPECTRAL PROPERTIES OF SYNCHROTRON RADIATION
\ssk
\ni 

The synchrotron radiation is emitted by e$^\pm$-pairs created in the process of
magnetic absorption of high-energy photons.
The character of the source function $Q_\pm$ of e$^\pm$-pairs depends primarily 
on the distribution of their pitch angles $\psi$
and on the richness of the cascades. 
In contrast to outer-gap models, small opening angles of magnetic field lines 
above the polar cap
result in confinement of pitch angles to a narrow range of low values.

In the case of $B_{\rm pc}= 10^{9}\G$
(Fig.1, upper panel) only one generation of e$^\pm$-pairs is produced.
Their source function $Q_\pm$ is almost monoenergetic and
the steady-state energy distribution of SR-cooled pairs 
($N_\pm \propto \gamma_\perp^{-2}$) results
in photon spectrum of SR with a well known power-law shape
$N_{\rm sr}(\epsilon) \propto \epsilon^{-{3 \over 2}}$. This spectrum extends
down to a low-energy turnover
$\epsilon_{\rm ct}$ 
determined by the condition
$\gamma_\perp \approx 1$
(O'Dell \& Sartori 1970). Below the turnover energy $\epsilon_{\rm ct}$, located around 
$0.1\keV$, 
the spectrum 
flattens, and may be described asymptotically as
$N_{\rm sr}(\epsilon) \propto \epsilon^{+1}$.
The SR component dominates over the CR component below $\sim 1\MeV$, i.e.
over the entire X-ray energy band.

In the case of $B_{\rm pc}= 10^{12}\G$ (Fig.1, lower panel) 
the SR component dominates over the CR in the energy range
of soft gamma-rays and hard X-rays.
The spectrum of the source function $Q_\pm$ of secondary pairs 
is spread over two decades in energy. Therefore, instead of
a single, well defined turnover energy,  
the SR spectrum reveals a gradual turnover which starts at $\sim 1\MeV$ 
(due to high values of $\gamma_\parallel$ as well as strong local $B$). At about $10\keV$ the
SR component falls below the level of the CR component.

Although for $10^9\G$ the spectrum of SR extends well into the soft X-ray band and 
dominates there,
its fractional contribution to the total radiation energy output is low.
It varies between $0.003$ (for $L_{\rm sd} = 10^{34}\ergs$) and 0.22 (for $10^{37}\ergs$).
For the case of 
$10^{12}\G$, the energy content of SR becomes significant for 
$L_{\rm sd} > 10^{33}\ergs$,
but the spectrum of SR is now confined to gamma-rays, and it turns over 
at $\lsim 1$ MeV.
In either case, the bulk of particle energy is converted into radiation
concentrated within the gamma-ray energy band.

\begin{figure}[t]
\begin{minipage}[l]{6cm}
\psfig{file=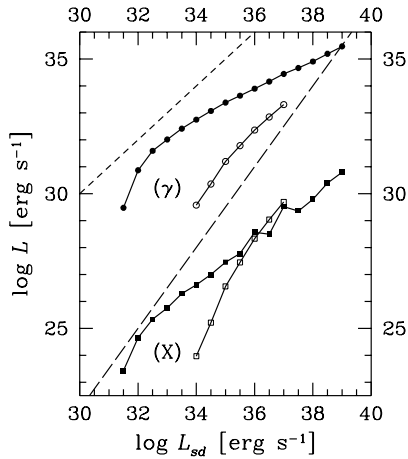, width=6cm}
\end{minipage}
\begin{minipage}[r]{6cm}
\caption{\ftn FIGURE 2. Evolution of high-energy, non-thermal luminosity across 
the spin-down luminosity space. 
The long upper curve connecting filled dots is the track
of gamma-ray luminosity $L_\gamma(>100\keV)$, and the long lower curve 
connecting filled squares is the track of X-ray luminosity $L_X(0.1 - 10\keV)$,
both calculated for a pulsar with $B_{\rm pc}=10^{12}\G$.
The short upper curve with open circles is 
for $L_\gamma$, and the short lower curve with
open squares is for $L_X$,
calculated for $B_{\rm pc}=10^{9}\G$.
The short-dashed line marks $L = L_{\rm sd}$.
The long-dashed line, marking the empirical relation of 
Saito et al.(1998) has been added for reference.
}
\end{minipage}
\end{figure}


\bsk
\ni 4. RELATIONS BETWEEN X-RAY AND $\gamma$-RAY LUMINOSITIES
\ssk
\ni 

The spectra calculated for a wide range of periods and magnetic
fields
show that the energy band of X-rays is never energetically important. 
Fig.2 presents X-ray and gamma-ray luminosities (\Lg~ and \Lx)
as a function of \edot, calculated by integrating the numerical spectra.
Predicted values of \Lx\t do not follow 
the empirical relation \Lx(\edot) found by Saito et al.(1988)
for pulsed components of X-ray emission. 
The calculated ratio \Lx/\Lg~ remains at the level between $10^{-6}$ 
and $10^{-4}$,
in disagreement with 
observations.
For the Crab, the observed ratio of \Lx/\Lg~ is about 0.47 (with \Lx~
and \Lg~ inferred for $\Omega_{\sss X} = \Omega_{\gamma} = 1$sr).
B1821$-$24, a `Crab-like' millisecond object, also 
poses a challenge to the polar-cap models.
Apart from the disagreement between the observed and predicted spectral 
slopes (photon index $\simeq 1.9$ (Saito et al. 1997) 
and $1.2$, respectively) the predicted ratio \Lx/\Lg 
$\simeq 2 \times 10^{-4}$ would make \Lg~ larger than \edot. 

On the other hand, the extremely low predicted ratios of \Lx/\Lg~ may offer
an explanation
for the case of B1706$-$44, a strong gamma-ray pulsar detected with {\it EGRET}
($L_\gamma = 2.5 \times 10^{34}\ergs$, Thompson et al. 1996) which shows
no pulsed X-ray emission. Its steady X-ray emission with
$L_X = 1.3\times 10^{33}\ergs$ (Becker et al. 1995), probably of nebular origin,
far exceeds the level of the predicted pulsed component: \Lx $\approx 3\times 10^{29}\ergs$.

\bsk
\baselineskip = 12pt
{\abstract \ni ACKNOWLEDGMENTS

This work has been financed by the KBN grant 2P03D-00911.
BR acknowledges travel grants 2P03C00511p01 and 2P03C00511p04.
}

\bsk
\baselineskip = 12pt


{\references \ni REFERENCES
\ssk

\ref Becker W., Brazier K.T.S., Tr{\"u}mper J., 1995, A\&A, 298, 528
\ref Becker W., Tr{\"u}mper J., 1997, A\&A, 326, 682 
\ref Daugherty J.K., Harding A.K., 1982, ApJ, 252, 337 
\ref Daugherty J.K., Harding A.K., 1996, ApJ, 458, 278 
\ref O'Dell S.L., Sartori L., 1970, ApJ, 161, L63 
\ref Rudak B., Dyks J., 1998a, MNRAS, 295, 337
\ref Rudak B., Dyks J., 1998b, MNRAS, submitted
\ref Saito Y., Kawai N., Kamae T., Shibata S., 1998,
in  Proc. of the International Conference on Neutron Stars and Pulsars,
eds. Shibazaki N., Kawai N. et al., Universal Academy Press, Tokyo, 295
\ref Saito Y. et al., 1997, ApJ, 477, L37
\ref Thompson D.J. et al., 1996, ApJ, 465, 385
}

}                      

\end{document}